\documentclass[preprint,12pt]{article}

 \usepackage{graphicx}

\usepackage{amssymb}
\usepackage{amsmath}

\begin{document}

\title{Self-organizing magnetic beads for biomedical applications\footnote{Accepted for publication in the \textit{Journal of Magnetism and Magnetic Materials.}}}

\author{Gusenbauer Markus\thanks{\texttt{Corresponding author, markus.gusenbauer@fhstp.ac.at}}, Kovacs Alexander, Reichel Franz, \\[0.1cm] Exl Lukas, Bance Simon, \"Ozelt Harald, Schrefl Thomas \\[0.1cm] University of Applied Sciences St. Poelten }

\maketitle

\begin{abstract}

In the field of biomedicine magnetic beads are used for drug delivery and to treat
hyperthermia. Here we propose to use self-organized bead structures to isolate circulating tumor cells using
lab-on-chip technologies. Typically blood flows past microposts functionalized with antibodies for circulating tumor
cells. Creating these microposts with interacting magnetic beads makes it possible to tune the geometry in size, 
position and shape. We developed a simulation tool that combines micromagnetics and discrete particle
dynamics, in order to design micropost arrays made of interacting beads. The simulation takes into account the
viscous drag of the blood flow, magnetostatic interactions between the magnetic beads and gradient forces from external aligned magnets. We developed a
particle-particle particle-mesh method for effective computation of the magnetic force and torque acting on the
particles. 

\end{abstract}






\section{Introduction}

\subsection{Circulating Tumor Cell (CTC)}

CTCs detach from a tumor and can remain in the blood even after the tumor is removed. Their presence increases the chance of net tumors 
developing. It is important to monitor the number of CTCs in the blood but their low concentration (a few per $\mu l$) when compared
to normal blood cells (around 5 million per $\mu l$) makes this difficult. A new and flexible method is required.\\

Lab-on-chips with fixed arrays are designed to use the properties of the CTCs to filter them. The most common ways are
mechanical filters or using antibodies. Membranes with slots smaller than tumor cells fulfill the requirements for a mechanical 
filter \cite{lu_parylene_2010}. Normal blood cells, they have similar sizes, are more deformable than CTCs and can go through. Another possibility 
is to use cylindrical posts coated with antibodies \cite{bell_isolation_2007}. If there is a large surface area the possibility 
for a connection with the tumor cells increases. \\

Recently a CTC-chip based on guided self-assembly of magnetic beads was proposed \cite{saliba_microfluidic_2010}. Saliba et al. use magnetic traps made by microcontact printing in order to 
create magnetic chains on a regular grid. They demonstrate cell capture using beads coated with antibodies. In this paper we show that tunable microfluidic chips
can be designed by carefully selecting the susceptibility and the applied magnetic fields. Using discrete particle simulations we compute the distance between particle
chains in a microfluidic chip as function of a homogenous magnetic bias field. Such devices might be useful to combine immunomagnetic with mechanical filtering
of CTCs \cite{maimonis_affinity_2010}. Maimonis et al. show that the capture efficiency can be increased in a microfluidic chip with 2 post arrays of different gap size which enables affinity
and size capturing. Our proposed chip technology may offer the possibility to switch between affinity and size capturing through changing the distance between 
antibody-coated chains by an external field.\\ 

\subsection{Content of the paper}

To optimize the design of CTC-chips there are several questions for us. Can we use self-organized magnetic beads to create filter-like structures? 
What types of particles are applicable? Properties to deal with are susceptibility and diameter of the beads. To answer this questions we developed a simulation model with interacting magnetic beads
to create particle chains. This is described in section 2 of the paper. In addition to the magnetic part we developed a blood cell model that flows through the channel with the chain barrier. The
described methods are used to demonstrate tunability of the chain gaps with an external magnetic field. Results are presented in section 3.     

\section{Methods}

\subsection{Magnetic particle dynamics}

\begin{figure}[h]
\begin{center}
 \includegraphics[scale=.3]{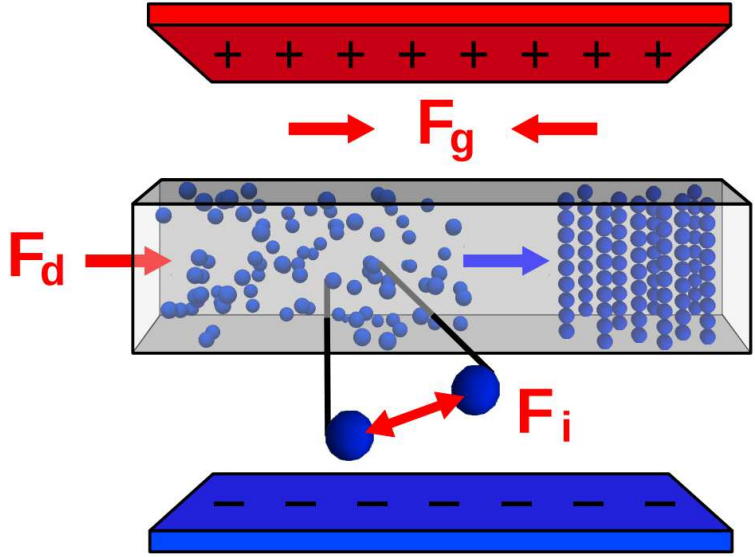}
\end{center}
 \caption{Formation of particle chains through the interaction of the drag force $F_d$ and magnetic forces. In the picture the 
magnetic forces are created by the magnetic field gradient $F_g$ and magnetic interactions $F_i$. The nonuniform magnetic field 
is created by 2 magnetic charge sheets. }
 \label{overview}
\end{figure}

In order to create a lab-on-chip device consisting of self-organizing micromagnetic beads we want to profit from an external magnetic field 
and the force of the blood flow. Fig. \ref{overview} shows the overview of the proposed microfluidic chip. Viscous blood flows into a pipe filled with soft-magnetic 
beads. Magnetic charge sheets with opposite magnetization orientation provide an external magnetic field.\\

Under the influence of the magnetic field a magnetic moment m is created in every particle. With the moments of two nearby beads and the distance r we got a 
formulation (Eqn. \ref{interaction}) of the interaction force $F_i$ for bead 2 and vice versa for bead 1 \cite{furlani_permanent_2001}. 

\begin{align}
\vec F_{1\rightarrow 2}= \frac{3\mu_0}{4\pi r^{5}}[(\vec m_1\vec r)\vec m_2 + (\vec m_2\vec r)\vec m_1 + (\vec m_1 \vec m_2)\vec r - \frac{5(\vec m_1\vec r)(\vec m_2\vec r)}{r^{2}}\vec r
\label{interaction}
\end{align}

The gradient force $\vec F_g$ (Eqn. \ref{Fstart}) on a bead is given by the negative gradient of the energy of the magnetic dipole moment $\vec m$ in the 
field $\vec B$.

\begin{align}
\vec F_g=\nabla(\vec m\vec B)
\label{Fstart}
\end{align}

In three dimensions it is given by 

\begin{align}
\vec F_g=\begin{pmatrix} F_x\\F_y\\F_z\end{pmatrix}=
	\begin{pmatrix} 
		    m_x\partial xB_x+m_y\partial xB_y+m_z\partial xB_z \\
		    m_x\partial xB_y+m_y\partial yB_y+m_z\partial yB_z \\
		    m_x\partial xB_z+m_y\partial zB_y+m_z\partial zB_z
	\end{pmatrix}
\label{Flong}
\end{align} 

We assume that in the x-y plane the external field is homogeneous due to the distance and size of the charge sheets. For the calculation 
of the gradient force only the z-field derived in y is important.  

\begin{align}
\vec F_g=\begin{pmatrix} 
		    0 \\
		    m_z\partial yB_z \\
		    0
	\end{pmatrix}
\label{Fend}
\end{align}

The blood flow creates a drag force \cite{mikkelsen_theoretical_2005} on the particles. At a low velocity of 460 $\mu m/s$ as used at most in 
CTC-chips \cite{bell_isolation_2007}, the flow is laminar. Turbulent flow would appear at Reynolds number (Eqn. \ref{Reynolds}) more than unity. 

\begin{align}
Re = \frac{2r\rho v}{\eta}
\label{Reynolds}
\end{align}

During laminar flow the Stokes Law (Eqn. \ref{Stokes}) is used to calculate the force $F_d$ 
on an object in a fluid, with particle radius r, density $\rho$ and viscosity $\eta$ of the fluid  and the relative velocity $v$ of the 
particle. Krishnamurthy et al. \cite{krishnamurthy_dynamics_2007} show that Stokes drag is a good approximation for the force acting on the beads.  

\begin{align}
F_d = 6\pi\eta rv
\label{Stokes}
\end{align}


For the above mentioned calculations of the magnetic particle dynamics we decided to expand the open-source particle simulator Yade \cite{kozicki_new_2008}. It 
provides for example gravity force and collision detection using various engines. Additionally we developed magnetic and fluidic engines. In the simulations we integrate 
the Newtons equation of motion for magnetic particles under the influence of the drag force of the fluid, the gradient force of the external magnetic field and the gradient force of 
the field created by the particles themselves.\\

For every particle in the pipe we calculate the magnetic field from the permanent magnets analytically \cite{akoun_3d_1984}. We represent
the 2 magnets by charge sheets as commonly done by magnetic simulation \cite{senanan_theoretical_2002}. The simulation 
parameters are listed in table \ref{parameters}.\\

For blood cell dynamics we implemented a cellular engine in Yade. This will be explained in the next section.\\

\begin{table}[h]
\begin{center}
\begin{tabular}{ | l | r | }
  \hline
  \textbf{Blood} & \\
  \hline                       
  velocity & $460\times10^{-6}$ m/s \\
  \hline 
  viscosity & 0.015 Ns/$m^2$ \\
  \hline 
  density & 1.055 g/$cm^3$ \\
  \hline
  \textbf{Particles} & \\
  \hline
  radius & $10^{-6}$ m\\
  \hline 
  susceptibility & 0.4 \\
  \hline 
  magnetic saturation & 1 T \\
  \hline 
  \textbf{Geometry} & \\
  \hline
  magnetic charge sheets & 10 x 10 mm \\
  \hline
  pipe & 10 x 0.05 x 0.02 mm \\
  \hline
  distance pipe-sheet & 2 mm \\
  \hline
  pole density sheets & 1.6 T \\
  \hline
\end{tabular}
\end{center}
 \caption{Simulation parameters}
\label{parameters}
\end{table}

\subsection{Blood cell dynamics}

\begin{figure}[h]
\begin{center}
   \includegraphics[scale=.55]{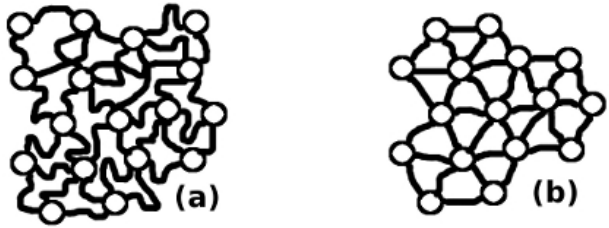}
\end{center}
  \caption{Cytoskeleton: a) relaxed, b) stretched}
  \label{cytosceleton}
 \end{figure}

The red blood cell consists of a cytoskeleton covered with a double layer membrane \cite{ikai_einfuehrung_2010}. It has a biconcave form in its native state with a diameter of 
around 8 $\mu m$, a surface of 135 $\mu m^{2}$ and a volume of 94 $\mu m^{3}$. The structure is built from 33000 hexagons which in reality would contain different proteins 
such as Spektrin and Aktin. In its loose state there is almost no interaction force between these proteins (Fig. \ref{cytosceleton}a). If there is an external force acting 
on the cell the cytoskeleton expands and gets stretched (Fig. \ref{cytosceleton}b). The membrane is an almost incompressible layer regarding the 
plane and shear stresses. Capillaries have a smaller diameter than the red blood cells, but because of their elastic properties they make their way using the form of a projectile. \\

\begin{figure}[h]
\begin{center}
\includegraphics[scale=.3]{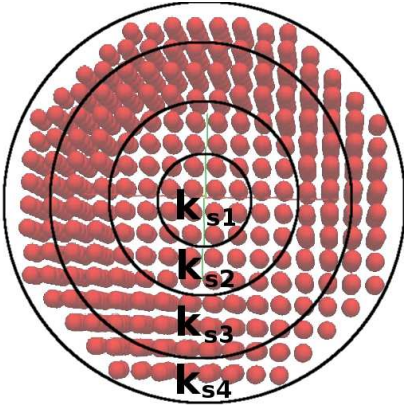}
\end{center}
 \caption{Nonlinear mass-spring model of red blood cell:  $k_{s1}$=500, $k_{s2}$=400, $k_{s3}$=300, $k_{s4}$=200 (Eqn. \ref{massSpring2})}
 \label{springStiff}
\end{figure}

Discrete element methods can be used to treat red blood cells numerically. In the following we will show that the behavior of red blood cells can be modeled using a nonlinear 
mass-spring-system (Fig. \ref{springStiff}). The model consists of nearly 600 spherical particles connected by springs arranged orthogonally. \\

The force between two particles i and j (Eqn. \ref{massSpring2}) depends on the respective spring constants, starting distance $l_0$ and the actual 
distance $d$. Yade provides the damping for all particles in the model. 


\begin{align}
 F(x_i,x_j)=k_s * \frac{d}{|d|} * (|d|-l_0) * |(|d|-l_0)|
\label{massSpring2}
\end{align} 

One possible way of characterizing a red blood cell is using an optical trap for extracting its shear modulus \cite{henon_new_????}. With an applied force on opposite 
sides of the cell, due to optical tweezers, the diameter is measured. From the approximated formula (Eqn. \ref{diameter}) the modulus decreases form the 
starting diameter $D_0$ with a function of the force F and the elasticity $\mu$.

\begin{align}
 D=D_0 - \frac{F}{2\pi\mu}
\label{diameter}
\end{align} 

\begin{figure}[h]
\begin{center}
  \includegraphics[scale=.45]{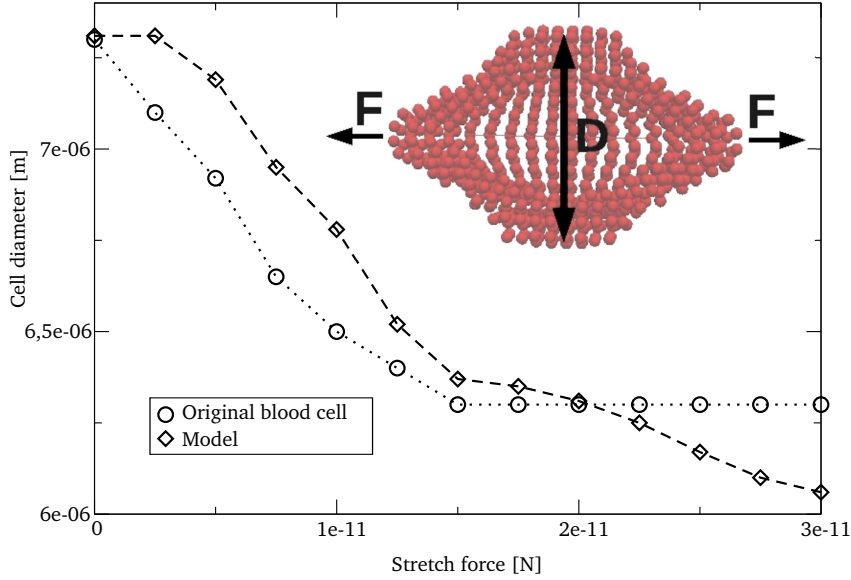}
\end{center}
 \caption{Stretch test of red blood cell, original \cite{henon_new_????} vs. model}
\label{stretchDia}
\end{figure}

To validate this model we computed this cell diameter as function of an applied stretch force (Fig. \ref{stretchDia}). Different spring constants are 
used to recreate a similar behavior in the stretching test. They decrease from the middle to the edge of the red blood cell (Fig. \ref{springStiff}). The modeled 
cell is accurate enough to make tests with the chain barrier. This will be demonstrated in the result section. \\

\section{Results}

\subsection{Single particle equilibrium position}

\begin{figure*}[h]

\begin{center}
 \includegraphics[scale=.45]{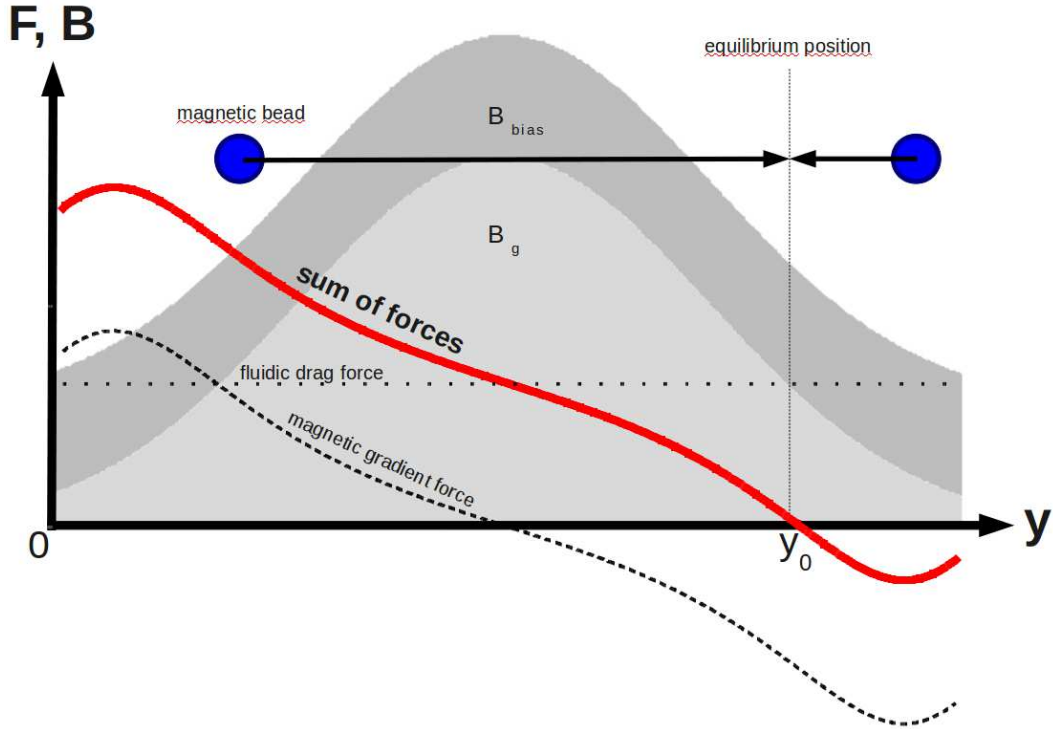}
 \caption{Magnetic charge sheets create a gradient field $B_g$. A homogenous bias field $B_g$ is added to the existing one for better tunability. Force balance of fluidic drag 
force $F_d$ and magnetic gradient force $F_g$ leading the particles to the equilibrium position $y_0$. }
\label{skizzeFgFd}
\end{center}
\end{figure*}

With a steady blood flow through the microfluidic chip the magnetic particles are flushed out without the presence of an external magnetic field. To find the equilibrium 
position $y_0$ of a single bead after applying the field we calculate the force balance of the drag force $F_d$ and the gradient force $F_g$ (Eqn. \ref{fg}). Fig. \ref{skizzeFgFd} 
shows this forces and their sum as function of the position. At the equilibrium state $F_d$ and $F_g$ cancel each other.\\ 

\begin{align}
 F_g(y)= F_d \quad\Longrightarrow\quad y_0
\label{fg}
\end{align} 

\begin{figure*}[h]

\begin{center}
 \includegraphics[scale=.4]{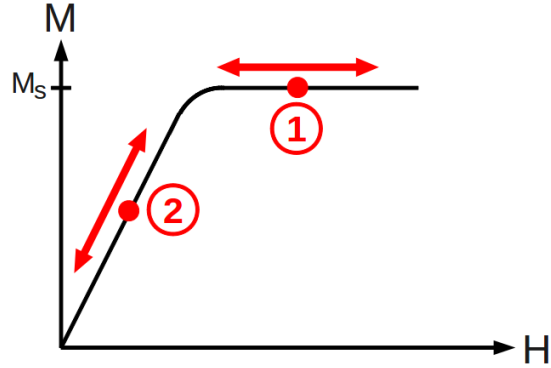}
 \caption{Operation points in the magnetization curve: 1. magnetic saturation, 2. initial magnetization (tunability) }
\label{magnCurve}
\end{center}
\end{figure*}

The magnetic moment $\vec m(y)$ of a single particle depends on its volume $V$, susceptibility $\chi$, which defines the ascending slope of the magnetization curve (Fig. \ref{magnCurve}), and 
the magnetic field. This field is the sum of the gradient field $B_g(y)$ and a homogenous bias field $B_{bias}$, which we will need to tune the device. 

\begin{align}
\vec m(y)=V \chi \frac{1}{\mu_0} (B_g(y) + B_{bias})   
\label{susc}
\end{align}

We have now 2 possible working points for the softmagnetic particles. In its saturated state (Fig. \ref{magnCurve}.1) the magnetic moment can't be further increased. For better tunability we are 
using the working point in the initial magnetization curve (Fig. \ref{magnCurve}.2). This will be shown in the next sections.

\begin{figure*}[h]

\begin{center}
 \includegraphics[scale=.5]{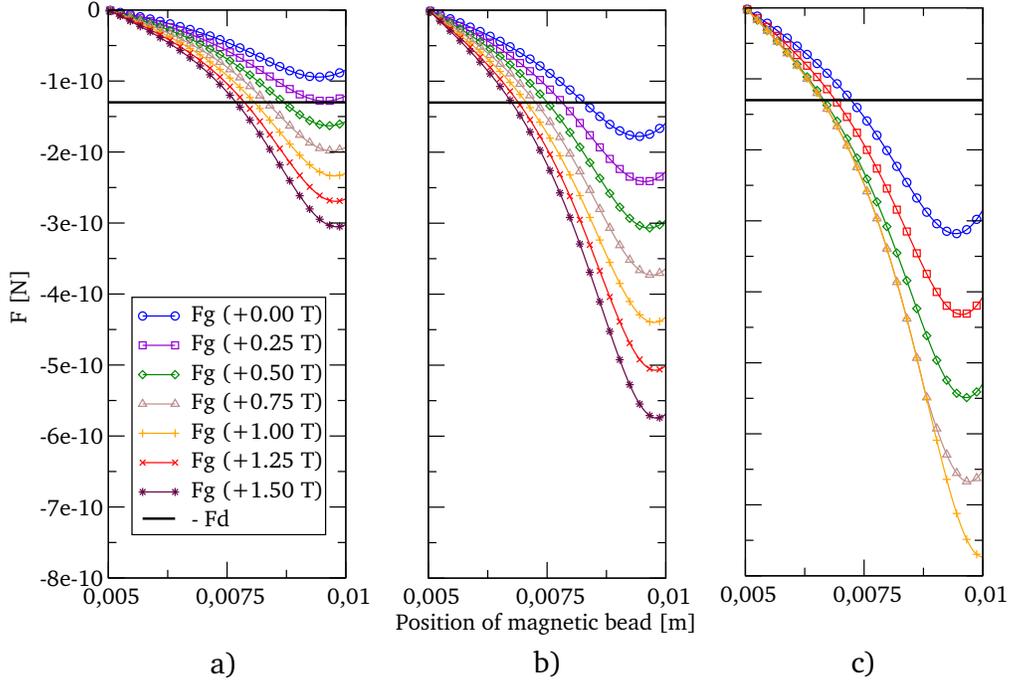}
 \caption{Gradient force $F_g$ as function of position with susceptibility a) 0.2 b) 0.4 c) 0.8. Crossing points with the drag force $F_d$ define the equilibrium position $y_0$ of a magnetic particle.}
\label{FgFd}
\end{center}

\end{figure*}

The maximum field from the charge sheets is in the center and decreases towards the edge (Fig. \ref{skizzeFgFd}). Fig. \ref{FgFd} shows the negative drag force and total magnetic force depending on
the position of the magnetic particles. The results show that the final position depends on the strength of the uniform bias field and the susceptibility $\chi$ of the particles. If there is no crossing point
the drag force is to high and the particles get washed out of the pipe (Fig. \ref{FgFd}a with a bias field of less than $0.5 T$).\\

If one of the simulation parameters changes, e.g. the viscosity of the blood, the drag force changes and therefore the external magnetic field has to change. Otherwise 
there probably wouldn't be a crossing point, and therefore no equilibrium position.

\subsection{Chain formation}

\begin{figure}[h]
\begin{center}
 \includegraphics[scale=0.50]{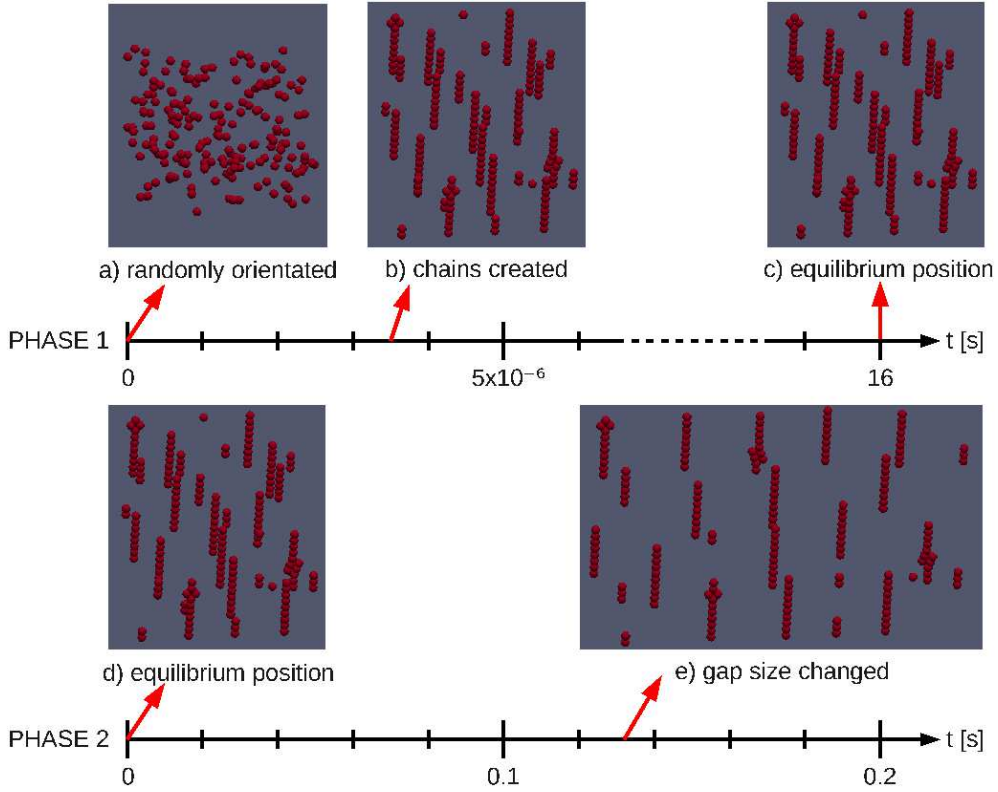}
\end{center}
 \caption{Timeline of the simulation: Phase 1 - Particle chain creation, Phase 2 - Changing the gap size}
\label{timeline}
\end{figure}

The simulation starts with softmagnetic beads of radius $1 \mu m$ and susceptibility $0.4$. They are randomly located in the fluidic channel (Fig. \ref{timeline}a). Immediately after applying the magnetic charge sheets the 
particles self-organize to chains according to the field lines (Fig. \ref{timeline}b). This chain formation is then shifted to the equilibrium position $y_0$ because of the force balance
explained in the section before. Because of chosen fluid velocity of 460 $\mu m/s$ this lasts around $16s$ (Fig. \ref{timeline}c).\\

After this procedure the filtering of the blood could start already. Different types of CTCs may need different distances between the chains for an optimal filtering. In the second phase of the 
simulation an additional homogenous bias field is applied on the chip. Depending on the field the chains change their gaps in around $0.13s$ (Fig. \ref{timeline}e). How it works will be explained in the next section.

\subsection{Tunable gap size}

\begin{figure}[h]
\begin{center}
 \includegraphics[scale=0.30]{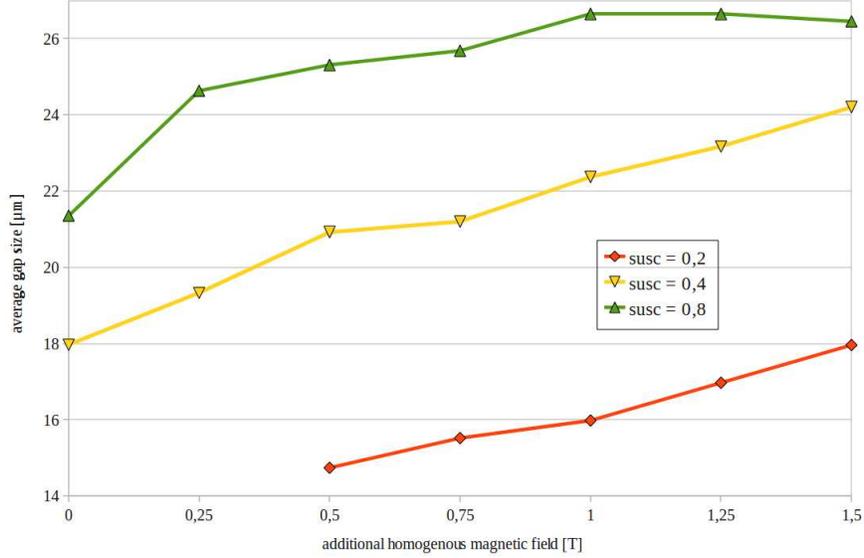}
\end{center}
 \caption{Average gap size between magnetic particle chains with different susceptibilities and additional external field}
\label{packLength}
\end{figure}

Eqn. \ref{susc2} shows the magnetic moment $\vec m(y)$ of a single bead with the existance of many particles. When we increase the bias field $B_{bias}$ the magnetic moment and 
therefore the magnetic field created in every particle gets more. But only if we are in the working point of the initial magnetization curve (Fig. \ref{magnCurve}.2). The interaction 
field $B_{i}$ is the sum of all other particle fields. This leads to a higher particle interaction force which causes a larger gap.\\

Fig. \ref{packLength} shows the increasing average gap size between the chains. In order to quantify this effect we computed the average distance between magnetic particle chains as a function of an additional homogenous 
magnetic field and susceptibility.

\begin{align}
\vec m(y)=V \chi \frac{1}{\mu_0} (B_g(y) + B_{bias} + B_{i})   
\label{susc2}
\end{align}

With a susceptibility of 0.2 resulting gaps occur only with an external field of more than $0.5 T$. In this case the force caused by the fluid is greater than 
the maximum force of the gradient. So the particles get washed out of the pipe. At an intermediate susceptibility of 0.4 the gap size increases linearly with the 
field from $18 \mu m$ to $24 \mu m$. At high susceptibility the total field will saturate the particles which leads to similar results after $0.5 T$. \\

\subsection{Particle density}

For the simulation result it is essential how dense the particles are filled into the pipe. For the particle density $\rho_{bead}$ we 
calculate the number of beads $N$ times the volume of one bead $V_{bead}$ over the volume of the wrapping box $V_{box}$ (Eqn. \ref{dens}).

\begin{align}
 \rho_{bead} = \frac{N * V_{bead}}{V_{box}}
\label{dens}
\end{align} 

\begin{figure}[h]
\begin{center}
 \includegraphics[scale=0.7]{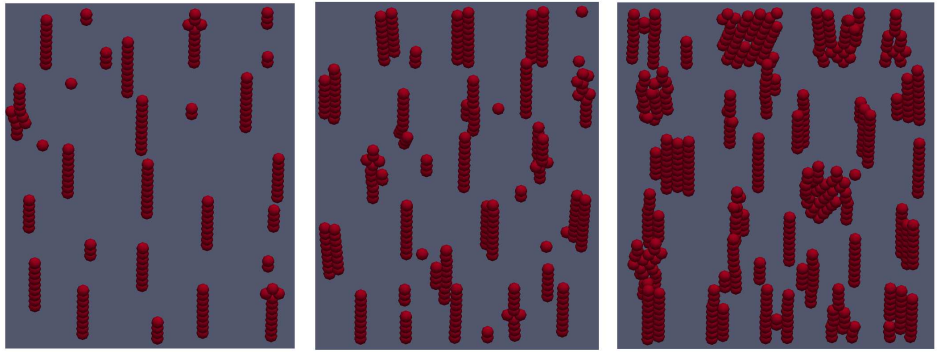}
\end{center}
 \caption{Endposition of chains with particle density 0.03, 0.07 and 0.11}
\label{densDiff}
\end{figure}

To evaluate the simulation we compared different starting densities and examined the resulting structures. Up to a density of 0.07 we find good results with not more than 2 bead 
thick chains. After that the particles get clumpy and can't be used for our application. Fig. \ref{densDiff} demonstrates the different results of bead densities.  


\subsection{Cell chain interaction}

\begin{figure}[h]
\begin{center}
 \includegraphics[scale=0.3]{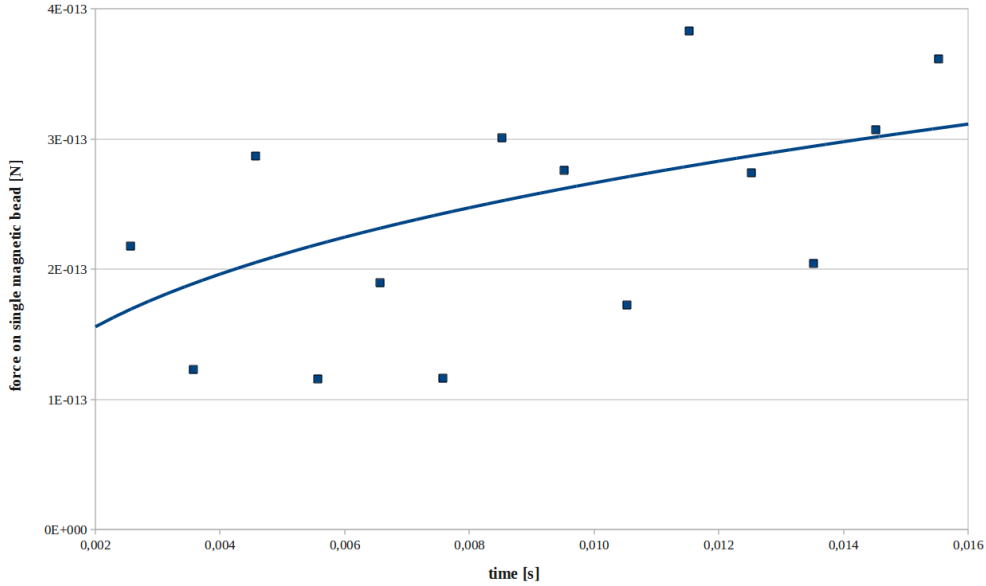}
\end{center}
 \caption{Force from the blood cell acting on a single magnetic bead of the chain structure.}
\label{cellForce}
\end{figure}

\begin{figure}[h]
\begin{center}
 \includegraphics[scale=1.0]{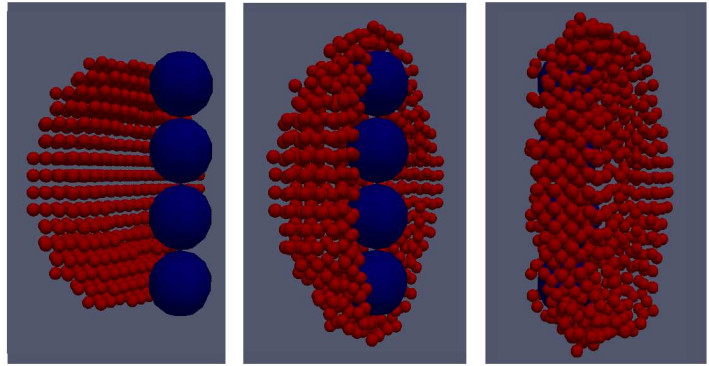}
\end{center}
 \caption{Red blood cell collides with a single magnetic chain for determing the contact force.}
\label{cellForcePic}
\end{figure}

Very important for the functionality of the filter is the stability of the chains. When a blood cell touches such a chain a certain force is applied on it. And it would destroy the structure
if it gets to high. To get a value of the force we started a simulation of our blood cell that flows directly on a particle chain that is fixed in space. Fig. \ref{cellForce} shows the average
force per $ms$ over time acting from the blood cell to the chain, especially on one magnetic bead of the chain. The simulation overview is shown in Fig. \ref{cellForcePic}. The magnitude of 
force on a single magnetic bead in the chain is 1000 times smaller than the drag force of the fluid or the gradient force of the magnetic charge sheets. So it doesn't influence the 
position and the stability of the chain structure. 

\section{Summary}

Flexible ways are important to get a high probability of catching cancer cells. Micromagnetic beads are established to create chain
barriers on demand. Gap sizes are modified as desired for different CTCs. Size-based filtration and the usage of antibodies covered on the beads 
can take place simultaneously. In this work we demostrated simulation tools for the design of micromagnetic CTC-chips.\\

First step of modelling blood cells was the creation of a nonlinear mass-spring-system. For the success of the project it is essential to create models of all 
kind of blood cells, healthy ones and circulating tumor cells. More complex fluidic behavior is also important for future work. Turbulences could occur due to the selfarrangement 
of the chains in the magnetic field or the interaction of magnetic beads and blood cells. \\

Results summarized:
\begin{itemize}
 \item Starting density of micromagnetic beads shouldn't be more then 0.07 for the ratio between volume of beads and wrapping box.
 \item Softmagnetic beads self-organizes to particle chains due to the external aligned magnetic charge sheets within a few $\mu s$.
 \item After about 16s the chain arrangement finds its equilibrium position with the force balance of fluidic drag force and magnetic forces.
 \item Applying an additional homogenous field changes the gap size between the chains according to different circulating tumor cells in around $0.13s$.
 \item A nonlinear mass spring cell model in the flow creates a force on a magnetic bead in a chain with a magnitude of 1000 smaller than the drag force or the gradient force. So it has no 
       influence on the chain structure.
\end{itemize}

\small{\textbf{Acknowledgment} The authors gratefully acknowledge the financial support
of Life Science Krems GmbH, the Research Association of Lower Austria.}\\




\newpage
\bibliographystyle{elsarticle-num}
\bibliography{biomed_app}







\end{document}